\documentstyle[prl,aps,epsfig,multicol,amssymb,amsfonts]{revtex}
\tighten
\renewcommand{\narrowtext}
{\begin{multicols}{2}\global\columnwidth20.5pc}
\renewcommand{\widetext}
{\end{multicols}\global\columnwidth42.5pc}
\input psfig

\begin{document}
\draft 
\title{Zero-frequency anomaly in quasiclassical ac transport:
Memory effects in a two-dimensional metal with a long-range random
potential or random magnetic field} 
\author{J.~Wilke,$^{1}$ A.~D.~Mirlin,$^{1,2,*}$
D.~G.~Polyakov,$^{2,\dagger}$ F.~Evers,$^1$ and P.~W\"olfle$^{1,2}$} 
\address{$^1$Institut f\"ur Theorie der Kondensierten Materie,
Universit\"at Karlsruhe, 76128 Karlsruhe, Germany}
\address{$^2$Institut
f\"ur Nanotechnologie, Forschungszentrum Karlsruhe, 76021 Karlsruhe,
Germany}
\maketitle
\begin{abstract}
We study the low-frequency behavior of the {\it ac} conductivity
$\sigma(\omega)$ of a two-dimensional fermion gas subject to a smooth
random potential (RP) or random magnetic field (RMF). We find a
non-analytic $\propto|\omega|$ correction to ${\rm Re}\,\sigma$, which
corresponds to a $1/t^2$ long-time tail in the velocity correlation
function. This contribution is induced by return processes neglected
in Boltzmann transport theory.  The prefactor of this $|\omega|$-term
is positive and proportional to $(d/l)^2$ for RP, while it is of
opposite sign and proportional to $d/l$ in the weak RMF case, where
$l$ is the mean free path and $d$ the disorder correlation
length. This non-analytic correction also exists in the strong RMF
regime, when the transport is of a percolating nature. The analytical
results are supported and complemented by numerical simulations.
\end{abstract}

\pacs{PACS numbers: 73.50.Bk, 05.60.Cd} 
\narrowtext

\section{Introduction}
\label{s1}

Within the conventional approach based on the Boltzmann equation, the 
{\it ac} conductivity of a two-dimensional electron gas (2DEG) is
described by the Drude formula
\begin{eqnarray}
&& \sigma_D(\omega)=\frac{\sigma_0}{1-i\omega\tau}\ , \label{drude}\\
&&\sigma_0=e^2\nu D\ ,\quad  D=\frac{v_{F}^2\tau}{2}\ , 
\label{sigma0}
\end{eqnarray}
where $\tau$ is the transport time, $\nu$ the density of states at the
Fermi level, $v_F$ the Fermi velocity, and $D$ the diffusion constant.
Equation (\ref{drude}) corresponds to an exponential falloff of the
velocity correlation function in the time representation,
\begin{equation}
\langle{\bf v}(t){\bf v}(0)\rangle=v_F^2e^{-t/\tau}\ ,\quad t>0~.
\label{markovv}
\end{equation}
This exponential behavior of $\langle{\bf v}(t){\bf v}(0)\rangle$
reflects the Markovian character of the Boltzmann equation description
and leads to analytical behavior of $\sigma_D(\omega)$ at $\omega\to
0$. It has been known for almost three decades, however, that these
features result from approximations made in the derivation of the
Boltzmann equation and are not generally shared by the exact solution
of the problem. More specifically, it was shown \cite{ernstwey} (see
\cite{hauge} for a review) that in the Lorentz gas model, where a
particle is scattered by randomly located hard discs of radius $a$ and
density $n_{s}$, there exists a ``long-time tail'' of the velocity
correlation function, which has the following form in two dimensions
(2D) in the limit $n_sa^2\ll 1$:
\begin{equation}
\langle{\bf v}(t){\bf v}(0)\rangle=-\frac{1}{4\pi n_s t^2}\ .
\label{lgasv}
\end{equation}
It leads to a  correction to the Drude conductivity, which is
non-analytic at $\omega\to 0$,  
\begin{equation}
\Delta{\rm Re}\,\sigma(\omega)= 
\sigma_0{1\over
8n_sl^2}|\omega|\tau=\sigma_0\frac{a}{3l}|\omega|\tau~, 
\quad
|\omega|\tau\ll 1\ ,
\label{longtimetail}
\end{equation}
where $l=v_F\tau$ is the mean free path and we substituted
$l=3/8n_sa$, the expression valid for the hard disc model.  We will
refer to this type of behavior of $\sigma(\omega)$ as a ``classical
zero-frequency anomaly''.  The long-time tail [which is of the form
$t^{-(d+2)/2}$ in $d$ dimensions] can be traced back to processes of
return of a particle to a region of extension $\sim l$ around the
starting point after moving diffusively during the time $t\gg\tau$
\cite{forfer}. These return processes give rise to non-Markovian
kinetics and are neglected in the Boltzmann equation.

After the discovery of weak localization the research interest has
shifted from the above (purely classical) effects to quantum
corrections to the conductivity. For a non-interacting 2D Fermi gas
the quantum (weak localization) correction is given by \cite{lr}
\begin{equation}
\Delta\sigma_{\rm wl}(\omega)=\sigma_{0}\frac{1}{\pi
k_{F}l}\ln{\left|\omega\tau\right|}\ ,
\label{weakloc}
\end{equation}
where $k_F$ is the Fermi wave vector. The weak localization
correction is of special interest, in particular, since it is
divergent at zero frequency, indicating the crossover to strong
localization. However, for weak disorder, $k_F l\gg 1$, the strong
localization is of purely academic interest, for its observation would
require exponentially small frequency and temperature and
exponentially large system size.

In recent years, there has been a revival of interest in semiclassical
transport properties of 2DEG.  This is motivated by the experimental
and technological importance of high mobility semiconductor
heterostructures, where impurities are located in a layer separated by
a large spacer $d\sim 100\, {\rm nm}$ from the 2DEG plane. The
(screened) random potential (RP) $V({\bf r})$ produced in the 2DEG
plane by the statistically distributed charged impurities (density
$n_i$) is characterized by the correlation function $W_{V}({\bf
r}-{\bf r}')=\langle V({\bf r})V({\bf r}') \rangle$, which has in
momentum space the form
\begin{equation}
\tilde W_{V}(q)=(\pi\hbar^2/m)^2 n_i e^{-2qd}\ ,
\label{wv}
\end{equation}
where $m$ is the particle mass. For $k_F d\gg 1$ (which is well
satisfied for the high-mobility samples), the potential varies
smoothly in space and can be treated in semiclassical terms. Such a
random potential is different from the Lorentz gas model in an
essential way. It is weak everywhere and shows close-to-Gaussian
fluctuations (since at $n_id^2\gg 1$ potentials produced by adjacent
scatterers strongly overlap), whereas in the Lorentz gas the potential
is zero outside and is infinite inside scatterers. Therefore, the
Lorentz gas results cannot be directly applied to the 2DEG, and the
problem has to be reconsidered for a realistic model of the random
potential.

Transport in a smoothly varying random magnetic field (RMF) is also of
major interest. One of the main motivations comes from the relevance
of this problem to the composite-fermion description of a 2DEG in a
strong magnetic field in the vicinity of half-filling of the lowest
Landau level ($\nu=1/2$) \cite{hlr}.  Exactly at $\nu=1/2$ the
composite fermions move in an effective magnetic field $B({\bf r})$
with zero average and impurity-induced spatial fluctuations
characterized by a correlation function $W_{B}({\bf r}-{\bf
r}')=\langle B({\bf r})B({\bf r}') \rangle$ of the form
\begin{equation}
\tilde W_{B}(q)=(2h c /e)^2 n_i e^{-2qd}\ .
\label{wb}
\end{equation}
A real long-range RMF can also be realized in semiconductor
heterostructures by attaching superconducting \cite{bending90,geim92}
or ferromagnetic \cite{mancoff95,ye96,rushforth99} overlayers or by
prepatterning of the sample (randomly curving the 2DEG layer)
\cite{gusev99}.  The strength of an RMF can be conveniently
characterized \cite{khvesh,mpw} by a dimensionless parameter
$\alpha=d/R_c^{0}$, where $R_c^{0}=v_{F}mc/eB_0$ is the cyclotron
radius in a typical field $B_{0}=\sqrt{\langle B^2\rangle}$ (magnitude
of the RMF fluctuations).  Within the composite-fermion theory of
\cite{hlr} this parameter is found to be equal to $1/\sqrt{2}$, if the
density of ionized impurities $n_i$ is assumed to be equal to the
electron density $n$ and if correlations between the impurity
positions are neglected. Experimental data for the magnetoresistivity
around $\nu=1/2$ are well described by the theory \cite{empw} with
somewhat smaller $\alpha\simeq 0.2 \div 0.35$ (the deviation can be
presumably attributed to the Coulomb correlations in positions of
impurities and other possible features related to technological
details of the sample preparation, as well as to approximations in the
composite-fermion theory).  We will concentrate in the main part of
this paper on the weak RMF case, $\alpha\ll 1$, when the transport is
of conventional diffusive nature.  The long-time tail in the case
$\alpha\gg 1$ (snake-state transport) will be discussed in
Sec.~\ref{s3.3}.

The following historical remark is in order here. After the initial
paper \cite{ernstwey} by Ernst and Weyland on the Lorentz gas model,
the $t^{-(d+2)/2}$ tail in the velocity correlation function of a gas
of particles scattered by static impurities has been discussed in a
number of publications, see in particular
\cite{maleev75,gantsevich81}. However, since there appear to be
neither a clear derivation nor explicit results for the long-time tail
in a smooth RP in the literature, we decided to present this material
in a self-contained form (Sec.~\ref{s2.1} and \ref{s3.1}). In fact,
our Eq.~(\ref{potcon1}) can be obtained from the mode-coupling
formalism of Ref.~\cite{belitz81}; however, the authors of that paper
concentrate on the critical regime of the metal-insulator transition
and do not consider the {\it ac} conductivity in the conducting phase
explicitly. As to the RMF problem, which constitutes the main focus of
the present paper, we are not aware of any treatment of the classical
non-analytic correction to the {\it ac} conductivity in the
literature.

For later use, we recall here the transport scattering rate entering
the Drude formula (\ref{drude}), (\ref{sigma0}), which, in the case of
weak long-range disorder, is found to be \cite{maw} 
\begin{eqnarray} \frac{1}{\tau}&=&\frac{1}{2\pi
m^2v_{F}^3}\int_0^{\infty}\!\!dq\;q^2 \tilde{W}_{V}(q) \quad
\mbox{(RP)}\ , \label{tauv} \\
\frac{1}{\tau}&=&\left(\frac{e}{mc}\right)^2\frac{1}{2\pi
v_{F}}\int_0^{\infty}\!\!dq\;\tilde{W}_{B}(q) \quad \mbox{(RMF)}\ .
\label{taub} \end{eqnarray}
Let us note that the Drude result is valid in the quantum regime as
well as in the classical limit. This is not, however, the case for
corrections to the Drude result and we will therefore consider both
the quantum theoretical treatment and purely classical description,
for different parameter ranges. 

We recall that the classical description of a quantum particle moving
in an RP or RMF characterized by a single spatial scale is a good
approximation if two conditions are satisfied: (i) the quantum
mechanical wavelength of the particle should be less than the
characteristic length $d$ of the disorder, i.e., $k_Fd\gg 1$; (ii) the
particle should move incoherently, i.e., the length over which it is
scattered out of its initial quantum state should be less than $d$:
$v_F\tau_s\ll d$, where $\tau_s$ is the single-particle life time. The
latter condition requires the random field to be sufficiently
strong. We will not address the regime $k_Fd\gg 1$, $v_F\tau_s\gg d$
in this paper. Our choice of models of disorder is motivated by actual
physical realizations and by considerations of calculational
feasibility. We will consider either a long-range RP or RMF governed
by Gaussian statistics. In the quantum mechanical calculation, we will
assume the transport scattering rate to be dominated by a white-noise
RP.

As discussed above in the context of the Lorentz gas, the fact that a
particle may revisit a given region of an RP/RMF after large time $t$
gives rise to a zero-frequency anomaly of the conductivity. Due to
this effect, the velocity correlation function acquires a power-law
behavior $\propto t^{-2}$ at long times, leading to ${\rm Re}\sigma
(\omega)\propto |\omega|$. The strength of this anomaly depends on the
probability of return into the region over which the RP or RMF is
correlated and may thus be expected to be proportional to a power of
$d/l$ (the power depends on the mechanism of scattering). It follows
that the magnitude of the classical zero-frequency anomaly for the
case of smooth disorder considered here may be much larger than the
weak localization correction (\ref{weakloc}) in a broad frequency
range. For a short-range potential, $k_Fd\alt 1$, the role of the
correlation length is played by the Fermi wavelength.

We turn now to the calculation of the low-frequency correction to the
Drude law induced by the return processes.

\section{ Return processes in the quantum-mechanical diagram technique}
\label{s2}

We begin by considering the case in which the leading contribution to
the transport scattering rate is given by a white-noise RP, while an
additional weak long-range RP or RMF induces correlations determining
the long-time tail. In this situation, a quantum-mechanical treatment
of the problem is appropriate. Apart from the theoretical convenience,
such a model with two types of disorder is also of experimental
relevance. Indeed, in essentially all realizations of {\it real} RMF
(as opposed to the fictitious RMF in the composite-fermion model) the
transport scattering rate is dominated by a random potential with a
relatively short correlation length (much shorter than that of RMF).

A white-noise RP is characterized by the correlation function
$\tilde{W}(q)=(2\pi\nu\tau_w)^{-1}$, where $1/\tau_w$ is the
corresponding scattering rate (we put here $\hbar=1$). We keep the
notation $1/\tau$ for the scattering rate related to a long-range RP
or RMF; $1/\tau\ll 1/\tau_w$. The diffusion process is described by a
sum of the ladder diagrams (a diffuson):
\begin{equation}
\Gamma({\bf q},\omega)=\frac{1}{2\pi\nu\tau_w^2}\frac{1}{Dq^2-i\omega}\ ,
\label{dif}
\end{equation}
where $D=v_F^2\tau_w/2$.

\subsection{Long-range random potential}
\label{s2.1}

\begin{figure}
\begin{center}
\includegraphics[clip]{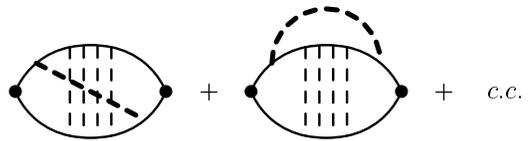}
\end{center}
\vspace{3mm}
\caption{Contribution to the conductivity due to return processes in
the presence of a long-range RP. The thin dashed lines (forming a
diffuson) correspond to a white-noise potential, while the thick
dashed line describes the long-range RP.}
\label{potdia}
\end{figure}

\begin{figure}
\begin{center}
\includegraphics[clip]{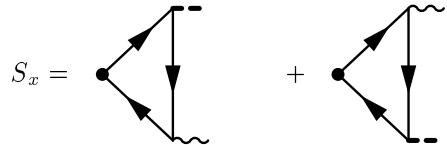}
\end{center}
\vspace{3mm}
\caption{Vertex parts of the diagrams shown in Fig.~\ref{potdia}. The
wavy line denotes the diffuson.}
\label{potvx}
\end{figure}

The contribution to the conductivity induced by
return processes is given by the sum of the diagrams shown in
Fig.~\ref{potdia}, yielding
\begin{equation}
\Delta\sigma=\frac{e^2}{2\pi}\int 
\!\frac{d^{2}q}{(2\pi)^2}\; S_{x}S_{x}\tilde{W}_{V}(q)\Gamma({\bf
q},\omega)\ .
\label{potcon}
\end{equation}
Here $S_x$ are the vertex parts which  
are represented in Fig.~\ref{potvx} and are given by the
following expression
\begin{eqnarray}
S_x&=&\int\!\frac{d^{2}p}{(2\pi)^2}\;
\frac{p_x}{m}G_{\epsilon_{F}}^R({\bf p})G_{\epsilon_{F}}^A({\bf
p}) \nonumber \\  
&&\quad \times \left(G_{\epsilon_{F}}^R({\bf p}-{\bf
q})+G_{\epsilon_{F}}^A({\bf p}+{\bf q})\right)\ ,
\label{potvxs}
\end{eqnarray}
where $G_{\epsilon_{F}}^R({\bf p})$ and  $G_{\epsilon_{F}}^A({\bf p})$
are the retarded and advanced Green's functions at the Fermi energy 
$\epsilon_F$.

The behavior of the correction $\Delta\sigma(\omega)$ at low $\omega$
is governed by small momenta, $q\sim (\omega/D)^{1/2}$, in the
integral (\ref{potcon}).  Therefore, we can make a small-$q$ expansion
of the vertex part (\ref{potvxs}). Expanding the integrand of
(\ref{potvxs}) up to terms linear in $q$ \cite{vw80}, we get
\begin{equation}
S_x({\bf q})=-iq_x\tau_w^2\ .
\label{potvx1}
\end{equation}
Note that a naive estimate of the linear-in-$q$ term would give
$S_x({\bf q}) \sim q_x \epsilon_F\tau_w^3$, but the two diagrams of
Fig.~\ref{potvx} cancel each other in this order and one has to go to
the next order in $1/\epsilon_F \tau_w$.  Substituting (\ref{potvx1}),
(\ref{dif}), (\ref{wv}) into Eq.~(\ref{potcon}), approximating the
correlation function $\tilde{W}_{V}(q)$ for small $q$ by its zero-$q$
value, and neglecting the $\omega$-independent part, we find the
following $\omega$-dependent contribution to the conductivity
\begin{equation}
\Delta\sigma(\omega)=\sigma_{0}\frac{\tilde{W}_{V}(0)}{4\epsilon_F^2
l_w^2}{\omega\tau_w\over i\pi}\ln(i\omega\tau_w)\ ,
 \quad |\omega|\tau_w\ll 1\ .
\label{potcon1c}
\end{equation}
The correction to the real part of the conductivity has therefore the
form
\begin{equation}
\Delta{\rm Re}\,\sigma(\omega)=\sigma_{0}\frac{\tilde{W}_{V}(0)}
{8\epsilon_F^2 l_w^2}|\omega|\tau_w~, \quad |\omega|\tau_w\ll 1\ .
\label{potcon1}
\end{equation}
The condition of validity $|\omega|\tau_w\ll 1$ given above
corresponds to the case $d\alt l_w$. In the opposite regime, $d\gg
l_w$, the formulas (\ref{potcon1c}), (\ref{potcon1}) still hold, but
the condition of their validity changes to $|\omega|\ll D/d^2$. The
same is valid for all the formulas for the non-analytic correction
that are given below.

We first consider the situation with only the white-noise potential
present, so that $\tilde{W}_{V}(q)=\tilde{W}_V(0)$. In this case, Eq.\
(\ref{potcon1}) yields \begin{equation} {\Delta {\rm Re}\,
\sigma(\omega)\over\sigma_{0}}= \frac{1}{2
(k_{F}l_w)^3}|\omega|\tau_w\ .  \label{wpotcon} \end{equation} We see
that the correction is small as $(k_{F}l_w)^{-3}$, i.e., much smaller
than the weak-localization correction (\ref{weakloc}), and is
therefore of minor interest. This conclusion changes, however, when we
return to the problem with the long-range potential (\ref{wv})
present. Equation (\ref{potcon1}) then gives 
\begin{equation}
{\Delta{\rm Re}\,\sigma(\omega)\over\sigma_{0}}=
4\pi\frac{\tau_w}{\tau}\left(\frac{d}{l_w}\right)^3\!|\omega|\tau_w\ .
\label{potconres} 
\end{equation} 
Now the correction does not contain the quantum small parameter
$(k_Fl_w)^{-1}$, which is replaced by the classical quantity
$d/l_w$. This prompts the expectation that the $|\omega|$ anomaly in
$\sigma (\omega)$ should be essentially a classical phenomenon. We
will demonstrate it explicitly in Sec.~\ref{s3} by calculating $\Delta
{\rm Re}\,\sigma (\omega)$ in the classical limit, where a long-range
RP constitutes the only type of disorder in the system. Note that the
classical limit requires two conditions to be met: $k_Fd\gg 1$ for all
relevant types of scatterers and also $\tilde{W}_{V}(0)\gg (\hbar
v_F)^2$ (the latter condition means smallness of the diffraction
smearing of a typical scattering angle; otherwise, it can be rewritten
as $v_F\tau_s\ll d$, where $\tau_s$ is the single-particle life time);
whereas Eq.~(\ref{potconres}) is obtained in the perturbative (Born)
limit $\tilde{W}_{V}(0)\ll (\hbar v_F)^2 $ under the condition that
the diffusion is due to short-range scatterers. It is also worth
mentioning here that, in view of $\tau\gg\tau_w$ and $v_F\tau\gg d$,
the correction (\ref{potconres}) is always small, $\Delta{\rm
Re}\,\sigma(\omega)/\sigma_{0}\ll 1$, in the range of its validity
[specified below Eq.~(\ref{potcon1})].

\subsection{Long-range random magnetic field}
\label{s2.2}

\begin{figure}
\begin{center}
\includegraphics[clip]{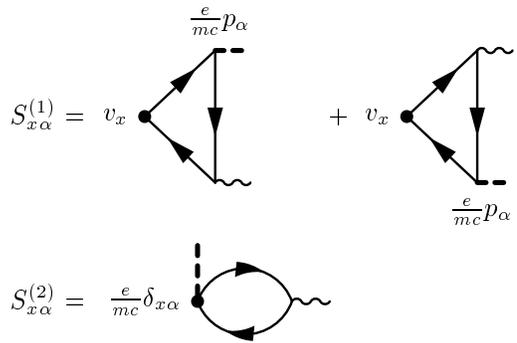}
\end{center}
\vspace{3mm}
\caption{Vertex parts of diagrams in a random magnetic field.}
\label{magvx}
\end{figure}

We consider now the same problem but with the long-range RP replaced
by a long-range RMF. Similarly to (\ref{potcon}), we have a
return-induced correction to the conductivity
\begin{eqnarray}
\Delta\sigma&=&\frac{e^2}{2\pi}\sum_{\alpha\beta}\int\!
\frac{d^{2}q}{(2\pi)^2}\;
\left(S_{x\alpha}^{(1)}+S_{x\alpha}^{(2)}\right)
\left(S_{x\beta}^{(1)}+S_{x\beta}^{(2)}\right)
\nonumber \\ 
&& \quad \times \langle A^{\alpha}({\bf q})A^{\beta}({\bf -q})\rangle 
\Gamma({\bf q},\omega)\ ,
\label{magcon}
\end{eqnarray}
where
\begin{equation}
\langle A^{\alpha}({\bf q})A^{\beta}(-{\bf q})
\rangle=\frac{\tilde{W}_{B}(q)}{q^2}
\left(\delta_{\alpha\beta}-\hat{q}_{\alpha}\hat{q}_{\beta}\right),
\quad \hat{q}_{\alpha}=\frac{q_{\alpha}}{|{\bf q}|}
\label{acor}
\end{equation}
is the vector potential correlation function. The vertex part
$S_{x\alpha}^{(1)}+S_{x\alpha}^{(2)}$ is now given by the sum of the
three diagrams shown in Fig.~\ref{magvx}.

Evaluating the vertex part at small $q$, we find that the diagrams
$S_{x\alpha}^{(1)}$ and $S_{x\alpha}^{(2)}$ cancel each other in the
order $q^0$ and the result is of the order of $q^2$:
\begin{equation}
S_{x\alpha}^{(1)}+S_{x\alpha}^{(2)}=
-\frac{e}{mc}q^2 \epsilon_{F}\tau_w^3\delta_{x\alpha}\ . 
\label{magvxs}
\end{equation}
Substituting this expression in (\ref{magcon}) and
neglecting an $\omega$-independent part, we find 
\begin{equation}
{\Delta{\rm Re}\,\sigma(\omega)\over
\sigma_{0}} = -\left(\frac{e}{mc}\right)^2
\frac{\tilde{W}_{B}(0)}{8v_{F}^2}|\omega|\tau_w~, 
\quad |\omega|\tau_w\ll1\ .
\label{magcon1}
\end{equation}
Using the explicit form of the correlation function (\ref{wb}), we get
\begin{equation}
{\Delta{\rm Re}\,\sigma(\omega)\over\sigma_{0}}
=-\frac{\pi}{2}\frac{d}{l}|\omega|\tau_w\ ,
\label{magconres}
\end{equation}
where $l=v_F\tau$ is the mean free path characterizing the RMF. We see
that the non-analytic conductivity correction (and correspondingly the
long-time tail of the velocity correlation function) has the opposite
sign as compared to the RP case, Eq.~(\ref{potconres}). This is a
general feature of the corrections induced by a weak long-range RMF,
as will be confirmed in Sec.~\ref{s3.2} by a classical calculation for
the case when such an RMF constitutes the only source of disorder.

\section{Purely long-range disorder: classical calculation of the
long-time tails}
\label{s3}

\begin{figure}
\begin{center}
\includegraphics[clip]{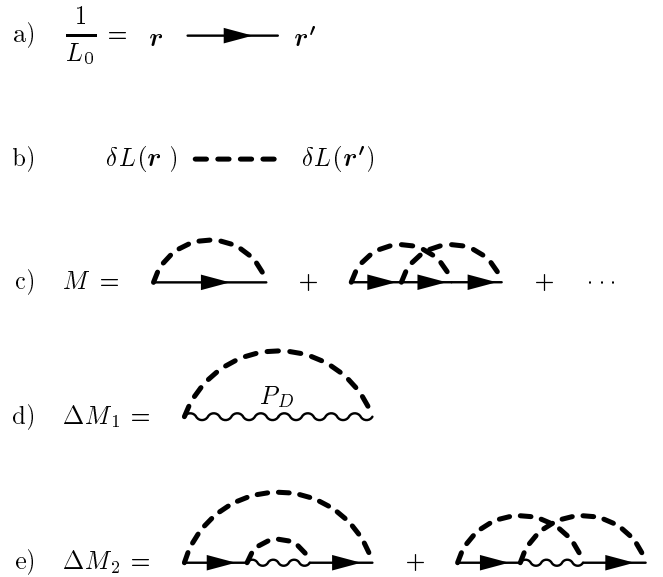}
\end{center}
\vspace{3mm}
\caption{Classical diagram technique: a) free propagator; b) disorder
correlation function; c) diagrammatic expansion for the memory
function $M$; d) first-order diagram for the memory function
representing a return process; wavy line corresponds to the diffusion
propagator $P_D$, Eq.~(\ref{pd}); e) second-order diagrams describing
return processes, which give the leading contribution to the
return-induced correction to $M$ in the random-potential case.}
\label{cdia}
\end{figure}

Having understood the nature of the $|\omega|$ anomaly at the level of
the quantum-mechanical diagram technique in the particular limit where
the transport scattering rate is dominated by a white-noise potential,
we turn to the case of purely long-range disorder (RP or RMF). In this
situation, the quantum-mechanical calculation is complicated and a
classical evaluation of the non-analytic correction is more
appropriate; it will also allow us to demonstrate explicitly that the
correction is of classical origin. We will employ a formalism similar
to the one used in \cite{mwepw} for the calculation of the
magnetoresistivity.  At the quasiclassical level, the fermion gas is
characterized by a distribution function $f(t,{\bf r},\phi)$, where
$\phi$ is the polar angle of the velocity. The equilibrium
distribution function is $f_0=\theta(\epsilon_F-\epsilon)$, where
$\theta$ is the step function. The deviation $\delta f(t,{\bf
r},\phi)$ from the equilibrium induced by an (infinitesimally small)
external electric field ${\bf E}e^{-i\omega t}$ has the form $\delta
f(t,{\bf r},\phi)=eEv_{F}\frac{\partial f_0}{\partial \epsilon}
e^{-i\omega t}g(\omega,{\bf r},\phi)$, with $g(\omega,{\bf r},\phi)$
obeying the Liouville equation
\begin{eqnarray}
(L_0+\delta L)g(\omega,{\bf r},\phi)&=&\cos{(\phi-\phi_E)}~;
\label{lio} \\
L_0&=&-i\omega+v_{F}{\bf n}\nabla~.
\end{eqnarray}
Here $\phi_E$ is the polar angle of the electric
field and ${\bf n}=(\cos{\phi}, \sin{\phi})$ the unit vector
determining the velocity direction. The term $L_0$ in the Liouville
operator corresponds to the free motion, while $\delta L$ describes
the disorder (RP or RMF). The current density is given by ${\bf
j}=-e\int\frac{d^2p}{(2\pi\hbar)^2} {\bf v} \delta f$, yielding
the longitudinal conductivity
\begin{equation}
\sigma(\omega)=e^{2}\nu v_{F}^2 \int\!\frac{d\phi}{2\pi}\;\left\langle
\cos{\phi} \frac{1}{L_0+\delta L} \cos{\phi}\right\rangle\ .
\label{ccon}
\end{equation}
Expanding (\ref{ccon}) in $\delta L$, averaging over the RP or RMF
(which is implicit in  $\delta L$), and resumming the series, we get
the {\it ac} conductivity in the form
\begin{equation}
\sigma(\omega)=\frac{\sigma_{0}/\tau}{-i\omega+M}\ .
\label{sigmaomega}
\end{equation}
Here $M$ is the self-energy (so-called memory function),
which can be conveniently represented
within a classical diagrammatic technique (similar to the one used in
Ref.~\cite{gantsevich81}), Fig.~\ref{cdia}. 
To leading order, $M$ is given by the first diagram of
Fig.~\ref{cdia}c, 
\begin{equation}
M_0=-2\int\!\frac{d\phi}{2\pi}\;\cos{\phi}\left\langle\delta
L\frac{1}{L_0}\delta L\right\rangle\cos\phi\ ,
\label{m0}
\end{equation}
reproducing the results (\ref{tauv}), (\ref{taub}) for the transport
scattering rate (see below) and, correspondingly, the Drude formula
(\ref{drude}). Corrections to the memory function $M(\omega)$, which
correspond to the return processes, are evaluated in Sec.~\ref{s3.1}
and \ref{s3.2} for the cases of RP and RMF, respectively.

\subsection{Long-range random potential}
\label{s3.1}

The fluctuating contribution to the Liouville operator due to the RP
is found to be \begin{equation} \delta L_{V}=\delta v({\bf r}){\bf
n}\nabla+(\nabla\delta v({\bf r})){\bf
n}_{\perp}\frac{\partial}{\partial \phi}\ , \label{lv} \end{equation}
where ${\bf n}_{\perp}={\bf \hat z}\times {\bf n}=(-\sin{\phi},
\cos{\phi})$ and $\delta v({\bf r})=v({\bf r})-v_F$ is the deviation
of the local velocity $v({\bf r})=\{(2/m)[\epsilon_F-V({\bf r})]\}^{1/2}$
from its average value $v_F$.  The leading-order contribution
(\ref{m0}) to the memory function reads 
\begin{eqnarray}
M_{0}&=&-\frac{2i}{p_{F}^2}\int\!\frac{d\phi}{2\pi}
\frac{d^2q}{(2\pi)^2}\;\sin{\phi}\sin{(\phi-\phi_q)}\nonumber\\
&\times&
\frac{q^2\tilde{W}_{V}(q)}{v_{F}q\cos{(\phi-\phi_{q})}-\omega-i0}
\sin{\phi}\sin{(\phi-\phi_q)} \ , \label{mccv} 
\end{eqnarray}
reproducing the transport scattering rate defined by Eq.~(\ref{tauv}),
$M_{0}=1/\tau$. The first-order diagram describing the return process
is represented in Fig.~\ref{cdia}d. The corresponding expression is
obtained by replacing the free propagator $1/L_0$ in (\ref{mccv}) by
the diffusion propagator
\begin{eqnarray} P_D({\bf
q},\phi,\phi')&=&{\gamma({\bf q},\phi)\gamma({\bf q},\phi') \over
Dq^2-i\omega}~, \label{pd} \\ \gamma({\bf q},\phi)&\simeq &1-iql\cos
(\phi-\phi_q)~,\quad ql\ll 1~, \nonumber 
\end{eqnarray} 
The replacement yields the return-induced first-order correction to
the memory function
\begin{eqnarray} \Delta M_1&=&\frac{2}{p_{F}^2}
\int\!\frac{d\phi}{2\pi}\frac{d\phi'}{2\pi}\frac{d^2q}{(2\pi)^2}\;
\sin{\phi}\sin{(\phi-\phi_q)} \nonumber \\ &\times&
q^2\tilde{W}_{V}(q)P_{D}({\bf q},\phi,\phi')
\sin{\phi'}\sin{(\phi'-\phi_q)}~.  \label{cmv} 
\end{eqnarray}
Evaluating the $\omega$ dependent part of (\ref{cmv}) at
$\omega\tau\ll1$ and approximating (as in the quantum-mechanical
calculation) $\tilde W_V(q)$ by its value at $q=0$, we find
\begin{equation} \label{adm1} {\Delta M_1(\omega)\over M_0}=
-\frac{\tilde{W}_{V}(0)} {16\epsilon_F^2 l^2} {\omega\tau\over
i\pi}\ln(i\omega\tau)~, 
\end{equation} 
which gives
\begin{equation}\label{estM1} {\Delta {\rm Re}\, M_1(\omega)\over
M_0}=-\pi\left({d\over l}\right)^3 |\omega|\tau~,  
\end{equation} for the specific form (\ref{wv}) of the correlator 
$\tilde{W}_V$.

Now let us show that, in actual fact, the leading contribution to the
non-Markovian correction to ${\rm Re}\, M(\omega)$ comes from
second-order processes described by the two diagrams in
Fig.~\ref{cdia}e, whereas that given by Eq.~(\ref{adm1}) can be
neglected in the first approximation. Specifically, the second-order
term $\Delta {\rm Re}\,M_2/M_0\sim (d/l)^2|\omega|\tau$ scales with a
smaller, as compared to $\Delta M_1$, power of the parameter $d/l\ll
1$, despite having one more impurity line. This, at first glance,
counterintuitive feature is related to the anomalous smallness of
$\Delta M_1$ in the otherwise ``regular" expansion in powers of $d/l$
(third- and higher-order terms in $\Delta M$ can be shown to be
negligible compared to $\Delta M_2$). We first explain this feature by
using the following power-counting argument. The $|\omega|$-anomaly in
$\Delta M$ comes from the integration over small $q$ of the form
$\int\!  d^2q\, q^2/(Dq^2-i\omega)$, where the numerator of the
integrand tends to zero as $q^2$ at $q\to 0$. In Eq.~(\ref{cmv}), the
factor of $q^2$ is related to the vanishing of the correlator
\begin{eqnarray*} \int d^2r \exp (-i{\bf qr})\left<\nabla\delta
v(0)\nabla\delta v ({\bf r})\right>\propto
q^2\tilde{W}_V(q)\end{eqnarray*} in the limit $q\to 0$, since the
correlator carries the small momentum $q$, the same as the diffuson,
according to Fig.~\ref{cdia}d. On the other hand, in second order in
$\tilde{W}_V$, the large momenta flowing through impurity lines are
``disentangled" from the small momentum $q$ carried by the
diffuson. The leading $q^2$-term comes now from the $O(ql)$
corrections to the diffusion propagator given by the factors
$\gamma({\bf q},\phi)$ in Eq.~(\ref{pd}). Let us count powers of $l$:
two factors $\gamma({\bf q},\phi)$ yield $q^2l^2$, whereas one loses
only $l^{-1}$ when going to the second order, which explains the total
gain of one power of $l/d$ as compared to Eq.~(\ref{estM1}).

The expression for $\Delta M_2$ at $\omega\to 0$ obtained from the sum
of the two diagrams in Fig.~\ref{cdia}e reads [we neglect the
dependence on $q$ everywhere but in $P_D({\bf q},\phi,\phi')$]
\begin{eqnarray} \Delta M_2&=&{4i\over p_F^4}\int\!{d\phi\over
2\pi}{d\phi'\over 2\pi}{d^2q\over (2\pi)^2}{d^2k\over
(2\pi)^2}k^4\tilde{W}_V^2(k)\nonumber \\ &\times&\cos\phi A({\bf
k},\phi)P_D({\bf q},\phi,\phi'){\rm Im}\,A({\bf k},\phi')\cos\phi'~,
\label{m2}\end{eqnarray}
where
\begin{eqnarray} A({\bf k},\phi)={\partial\over\partial\phi}{\sin^2
(\phi-\phi_k)\over v_Fk\cos
(\phi-\phi_k)-i0}{\partial\over\partial\phi}~.\end{eqnarray}
We thus obtain
\begin{eqnarray} {\Delta {\rm Re}\,M_2(\omega)\over
M_0}=-{|\omega|\tau\over 32\epsilon_F^4}\int\!{d^2k\over
(2\pi)^2}k^2\tilde{W}_V^2(k)~.  \label{mfrpclass} \end{eqnarray}
Since
\begin{equation}
\label{adm2}
{\Delta{\rm Re}\,\sigma(\omega)\over\sigma_0}\simeq -
{\Delta {\rm Re}\,M(\omega)\over M_0} \equiv
{\Delta{\rm Re}\,\rho(\omega)\over\rho_0}\ ,
\end{equation}
where $\rho(\omega)=\sigma^{-1}(\omega)$ is the {\it ac} resistivity,
we get finally, using Eq.\ (\ref{wv}) for $\tilde{W}_V$,
\begin{equation}\label{rpclass}
{\Delta {\rm Re}\,\sigma (\omega)\over \sigma_0}={3\pi\over
8}\left({d\over l} \right)^2|\omega|\tau~, \quad |\omega|\tau\ll 1~.  
\end{equation}
The prefactor of the $|\omega|$-correction to ${\rm Re}\,\sigma
(\omega)$ is positive, as in the quantum-mechanical result
(\ref{potconres}) and in the Lorentz gas formula
(\ref{longtimetail}). Note that the correction (\ref{rpclass}) matches
that for the Lorentz gas (\ref{longtimetail}) at $n_sd^2\sim 1$, as
expected---since this condition separates two extremes of strongly
non-Gaussian (Lorentz gas) and Gaussian [Eq.~(\ref{wv})] disorder. On
the other hand, the crossover between Eqs.~(\ref{rpclass}) and
(\ref{potconres}) occurs when the two following conditions are
fulfilled: $k_Fd\sim 1$ and $\tilde{W}_V(0)\sim (\hbar v_F)^2$ [cf.\
the definition of the classical limit for Gaussian disorder after Eq.\
(\ref{potconres})].

\subsection{Long-range random magnetic field}
\label{s3.2}

The fluctuating contribution to the Liouville operator induced by the RMF
has the form
\begin{equation}
\delta L_{B}=\frac{e}{mc}B({\bf r})\frac{\partial}{\partial\phi}\ .
\label{lb}
\end{equation}
The lowest-order contribution (\ref{m0}) to the  memory function
\begin{eqnarray}
M_{0}&=&-2i\left(\frac{e}{mc}\right)^2\int\!
\frac{d\phi}{2\pi}\frac{d^2q}{(2\pi)^2} \sin{\phi} \nonumber \\ 
&&\quad \times \frac{\tilde{W}_{B}(q)}
{v_{F}q\cos{(\phi-\phi_{q})}-\omega-i0} \sin{\phi}
\label{mb0}
\end{eqnarray}
reproduces again the corresponding transport scattering rate 
(\ref{taub}),  $M_{0}=1/\tau$. The first-order correction due to return
processes, Fig.~\ref{cdia}d, reads 
\begin{eqnarray}
\Delta M_1&=&2\left(\frac{e}{mc}\right)^2
\int\!\frac{d\phi}{2\pi}\frac{d\phi'}{2\pi}\frac{d^2q}{(2\pi)^2}\;
\sin{\phi} \nonumber \\
&&\quad\times \tilde{W}_{B}(q) P_{D}({\bf q},\phi,\phi')\sin{\phi'}\ ,
\label{cmb}
\end{eqnarray}
where the factors $\gamma({\bf q},\phi)$ should be included in $P_{D}({\bf
q},\phi,\phi')$, which gives
\begin{equation}
{\Delta{\rm Re}\,\sigma(\omega)\over\sigma_{0}}=
-{\Delta {\rm Re}\,M(\omega)\over M_0}=
-\left(\frac{e}{mc}\right)^2
\frac{\tilde{W}_{B}(0)}{8v_{F}^2}|\omega|\tau\ .
\label{cmagcon}
\end{equation}
Note that, in contrast to the case of RP, the leading contribution to
the return-induced correction $\Delta M$ in RMF comes from the
first-order processes. This is because the RMF scattering operator
$\delta L_B$ (in contrast to its RP counterpart $\delta L_V$) does not
involve spatial gradients. Using the RMF correlation function
(\ref{wb}), we finally get
\begin{equation}
{\Delta{\rm Re}\,\sigma(\omega)\over\sigma_{0}}=
-\frac{\pi}{2}\frac{d}{l}|\omega|\tau
=-\pi\alpha^2|\omega|\tau\ ,\quad |\omega|\tau\ll 1.
\label{cmagres}
\end{equation}

We have found, therefore, in agreement with the quantum-mechanical
result (\ref{magconres}), a negative sign of the
$|\omega|$-contribution to the conductivity. Analyzing the
calculation, one can trace the difference in sign [as compared to the
Lorentz gas result (\ref{longtimetail}) and the RP results
(\ref{potconres}), (\ref{rpclass})] back to the fact that the RMF
scattering operator (\ref{lb}) is odd with respect to time reversal. 

\begin{figure}
\includegraphics[width=0.9\columnwidth,clip]{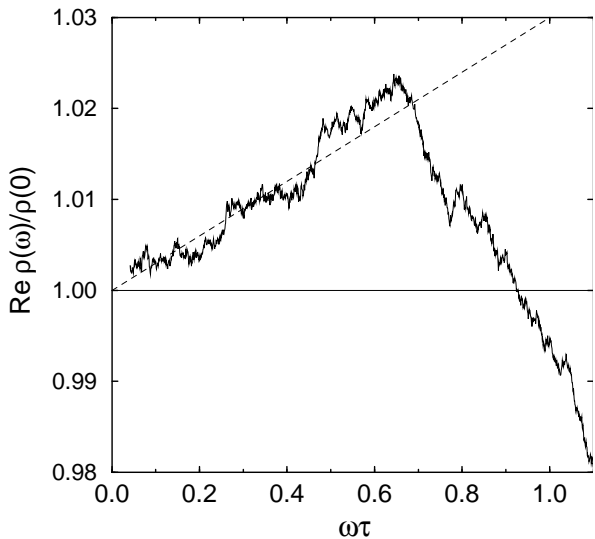}
\vspace{3mm}
\caption{Real part of the {\it ac} resistivity in the RMF with
$\alpha=0.5$ normalized to its $\omega=0$ value. The dashed line is a
guide for the eye, $\Delta{\rm Re}\,\rho(\omega)/\rho(0)\propto |\omega|$. 
The positive prefactor of the 
$|\omega|$-correction to ${\rm Re}\,\rho$ corresponds to a
negative prefactor for ${\rm Re}\,\sigma$. }
\label{1otau}
\end{figure}

To check the above analytic findings, we have performed numerical
simulations of the classical motion of a particle in an RMF. The
results obtained for the memory function \cite{foot2} at $\alpha=0.5$
are shown in Fig.~\ref{1otau}. We find a positive
$|\omega|$-correction to the real part of the memory function (or,
equivalently, resistivity), which corresponds to a negative correction
to ${\rm Re}\,\sigma$, in agreement with the theoretical result
(\ref{cmagres}). The magnitude of the correction is, however,
considerably smaller than Eq.~(\ref{cmagres}) would predict. We
attribute this discrepancy to the fact that Eq.~(\ref{cmagres}) was
derived for $\alpha\ll 1$ and, apparently, the numerical value of the
coefficient in this formula cannot be trusted for $\alpha$ as large as
0.5 \cite{foot3}. Unfortunately, at smaller values of $\alpha\alt
0.2$, the effect becomes so weak that it is swamped by the
statistical noise. A smaller value of the coefficient at $\alpha=0.5$
[as compared to the $\alpha\ll 1$ formula (\ref{cmagres})] is further
consistent with the fact that at $\alpha\gtrsim 1$ the coefficient
changes sign and the correction to the conductivity becomes positive
(see below). Let us also note that the range of validity of the
$|\omega|$-correction found numerically is in full agreement with the
theoretical expectation ($\omega\tau\lesssim 1$). Indeed, as is seen
in Fig.~\ref{1otau}, the linear increase of ${\rm Re}\,\rho(\omega)$
holds up to $\omega\tau\simeq 0.65$ where it transforms (rather
abruptly) into a falloff (related to the ballistic motion on time
scales $t\lesssim\tau$).

\subsection{Strong random magnetic field: long-time tail in 
transport on a percolating network}
\label{s3.3}

In a strong RMF ($\alpha\gg 1$) the character of the transport changes
drastically. In this regime, the diffusion takes place in a restricted
space and is determined by a small fraction of
trajectories---so-called ``snake states''
\cite{mueller92,chklov93}---which wind around the $B({\bf r})=0$
contours. Since the snake states can go over from one $B({\bf r})=0$
line to another at saddle points of the RMF (where the two contours
come sufficiently close to each other), they propagate effectively on
a percolating network \cite{lee94,mpw,empw} for which such saddle
points serve as nodes. This network is characterized \cite{empw} by a
typical length of a link, $L_s\sim d\alpha^{14/9}$, and a typical
distance between two neighboring saddle-points (size of an elementary
cell), $\xi_s\sim d\alpha^{8/9}$. The different scaling of $L_s$ and
$\xi_s$ with $\alpha$ is due to the fact that the structure of the
links of the network is fractal. The network is chiral, i.e., the
links are directed; each node has two incoming and two outgoing links.
Since the snake-state velocity is of the order of the Fermi velocity,
a characteristic time of traversal of a link is $\tau_s\sim L_s/v_F$.
The quasiclassical {\it dc} conductivity in this regime was calculated
in \cite{mpw,empw}, the result being $\sigma\sim k_F
d/\alpha^{1/2}\ln^{1/4}\alpha$.

In Ref.~\cite{empw} we argued, on phenomenological grounds, that for
such a percolation-type transport problem there should be a
non-analytic contribution to the {\it ac} conductivity of the form
\begin{equation}
\frac{\Delta{\rm Re}\,\sigma(\omega)}{\sigma(0)}\sim|\omega|\tau_{s}, 
\quad |\omega|\tau_s\ll 1\ .
\label{largea}
\end{equation}
Below we demonstrate how this result comes about in a network model
due to fluctuations in the geometry of the network.

\begin{figure}
\begin{center}
\includegraphics[clip]{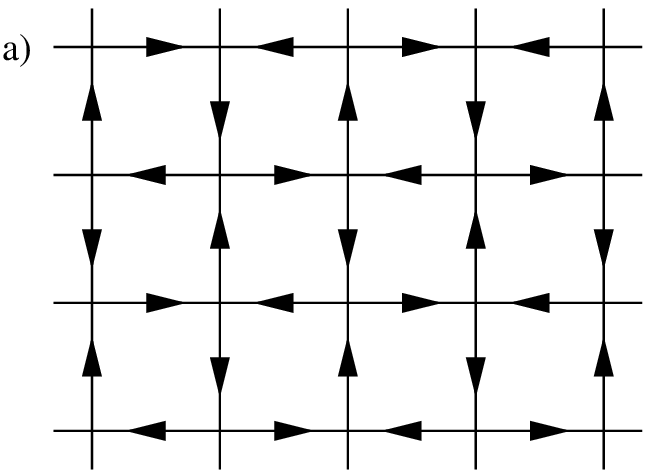}

\vspace{8mm}

\includegraphics[clip]{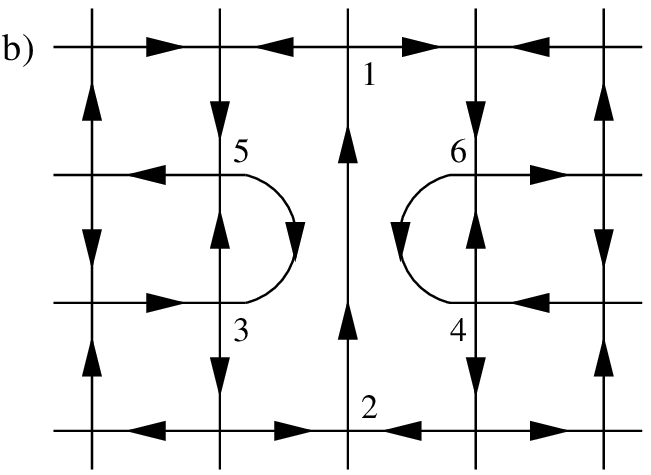}
\end{center}
\vspace{3mm}
\caption{Chiral network model: a) regular network; b) defect on the 
lattice.}
\label{net}
\end{figure}

Let us start with a regular square network (Fig.~\ref{net}a) with all
links characterized by the same distance $\xi_s$ between the end
points and by the same ``flight time'' $\tau_s$.  We also assume that
the probability of turning in either of two allowed directions at each
node is $1/2$. The classical diffusion constant is then
$D=\xi_s^2/4\tau_s$. It is straightforward to see that there is no
memory effect in classical transport on the regular network: the
velocity correlation function is exactly zero for $t>\tau_s$.  Let us
now study the effect of fluctuations in the network geometry, i.e., in
vectors connecting the beginning and the end of individual
links. Since to describe such fluctuations quantitatively in a real
percolating network is hardly possible, we consider the following
model. We imagine the regular lattice considered above perturbed by a
small fraction $n_d\ll 1$ of ``defects'' of the type shown in
Fig.~\ref{net}b (a defect can have any of four possible
directions). We assume the flight times of all links to be equal (we
will discuss the effect of fluctuations in the flight times later).

For each lattice site $j$, we label adjacent links as $(j\mu)$, with
$\mu=1,2$ for incoming and $\mu=3,4$ for outgoing links. The
velocity-velocity correlation function for a time $t=n\tau_s$ (with an
integer $n$) can be written as \begin{equation} \label{net1} \langle
{\bf v}(n\tau_s){\bf v}(0)\rangle={1\over 4N}\sum_{ij} \sum_{\mu=1,2}
\sum_{\nu=3,4}{\bf v}_{i\mu}{\bf v}_{j\nu} P_{ij}((n-1)\tau_s)\ ,
\end{equation} where $N$ is the normalization factor (total number of
sites), $P_{ij}(t)$ is the probability of moving from a site $i$ to a
site $j$ in a time $t$, and ${\bf v}_{i\mu}=\bbox{\xi}_{i\mu}/\tau_s$
is the velocity at the link $(i\mu)$. The majority of the sites $i,j$
will give zero contribution to (\ref{net1}) after the summation over
$\mu$ and $\nu$, since the velocities of the two outgoing (or two
incoming) links are exactly opposite to each other for the regular
lattice. A non-trivial contribution will come from terms with both $i$
and $j$ lying at a defect. Indeed, consider the term
with $i=1$, $j=2$ (Fig.~\ref{net}b). The corresponding contribution to
(\ref{net1}) is
\begin{equation}
\label{net2}
n_dP_{12}(t-\tau_s)\left({\xi_s\over\tau_s}\right)^2\ .
\end{equation}
The probability density in a continuum model for a
diffusing particle to move a distance ${\bf r}$ in a time $t$ is
\begin{equation}
\label{adm3}
P(t,{\bf r})={1\over 4\pi Dt}e^{-{\bf r}^2/4Dt}\ .
\end{equation}
Therefore, the probability $P_{12}(t)$ for $t\gg\tau_s$ is
\begin{equation}
\label{adm4}
P_{12}(t)= {\tau_s\over\pi t}\left[1+O\left({\tau_s\over t}\right)\right]\ .
\end{equation}
This return process yields a contribution
to the velocity correlation function of the form 
$$
n_d\left({\xi_s\over\tau_s}\right)^2{\tau_s\over \pi t}\propto {1\over t}\ .
$$
However, this $1/t$-contribution is canceled if we take into account
the terms with $i=3,4$ and $j=5,6$ as well. The total contribution
reads
\begin{eqnarray}
\label{net3}
&&\langle{\bf v}(t){\bf v}(0)\rangle= n_d\left({\xi_s\over\tau_s}\right)^2
[P_{12}(t-\tau_s)-P_{15}(t-\tau_s) \nonumber \\
&&-P_{32}(t-\tau_s)+
{1\over 2}P_{35}(t-\tau_s)+{1\over 2}P_{36}(t-\tau_s)]\ .
\end{eqnarray}
Since all of the relevant return probabilities $P_{ij}$ have the form 
(\ref{adm4}), the $1/t$ terms cancel. It is easy to see that this
cancellation has a general character, i.e., is independent of the
particular structure of the defect. We thus conclude that the result
is of the next  order in $\tau_s/t$, 
\begin{equation}
\label{adm5}
\langle {\bf v}(t){\bf v}(0)\rangle \sim - n_d
\left({\xi_s\over\tau_s}\right)^2  \left({\tau_s\over t}\right)^2\ .
\end{equation}
While we do not calculate the numerical coefficient in (\ref{adm5})
\cite{aside}, we see no reason which would require it to be zero, so
that we believe that it is generically non-zero.  Setting now $n_d\sim
1$ for a realistic (strongly fluctuating) network results in a
non-analytic correction to the conductivity of the form
(\ref{largea}). As to the sign of the effect, we have to resort to
numerical simulations (see below). 

Let us now consider the effect of fluctuations in flight time. We 
return to the regular square lattice (with the lattice constant
$\xi_s$), but now allow for variation of the flight times $\tau_\mu$
from one link to another. We will show that in this model the $1/t^2$
tail does not exist.  Equation (\ref{net1}) for the velocity-velocity
correlation function is now modified as follows
\begin{eqnarray}
\label{net4}
&&\langle{\bf v}(t){\bf v}(0)\rangle={1\over 4N\tau_s}
\sum_{ij}\sum_{\mu=1,2}\sum_{\nu=3,4}\nonumber\\
&&\times\left\langle
\int_0^{\tau_{i\mu}}d\tau\int_0^{\tau_{j\nu}}d\tau'
{\bf v}_{i\mu}{\bf v}_{j\nu} P_{ij}(t-\tau-\tau')\right\rangle\ ,
\end{eqnarray}
where ${\bf v}_{i\mu}=\bbox{\xi}_{i\mu}/\tau_\mu$ and 
$\tau_s=\langle\tau_\mu\rangle$. Since the fluctuations of the flight
times of different links are uncorrelated, a non-zero contribution to
(\ref{net4}) comes only from neighboring sites $i,j$ connected by
a link going from $j$ to $i$ [in other words, one of the links $(i\mu)$
should be identical to one of the links $(j\nu)$]. We find thus
\begin{eqnarray}
\label{net5}
&&\langle{\bf v}(t){\bf v}(0)\rangle={1\over 2}n_d{\xi_s^2\over\tau_s}
\nonumber\\
&&\times\left\{\left\langle\int_0^{\tau_1}d\tau
\int_0^{\tau_2}d\tau'{1\over
\tau_1\tau_2} P_{ij}(t-\tau-\tau')\right\rangle_{\tau_1,\tau_2}
\right. \nonumber \\
&&-\left.\left\langle\int_0^{\tau_1}d\tau\int_0^{\tau_1}d\tau'{1\over
\tau_1^2} P_{ij}(t-\tau-\tau')\right\rangle_{\tau_1}\right\}\ .
\end{eqnarray}
Expanding $P_{ij}(t-\tau-\tau')\simeq 1/\pi(t-\tau-\tau')$ in $\tau$
and $\tau'$, we see that the terms of the $1/t$ and $1/t^2$ orders
cancel, and the leading non-vanishing contribution is of the $1/t^3$
order, so that the corresponding contribution to ${\rm
Re}\,\sigma(\omega)$ shows a weak non-analyticity $\propto
\omega^2\ln|\omega|$ only. Note that for a non-directed network,
fluctuations of the flight time yield a still weaker non-analyticity 
$\Delta {\rm Re}\,\sigma(\omega)\propto |\omega|^3$, or, equivalently,
a $1/t^4$ long-time tail \cite{jw}.
 
\begin{figure}
\includegraphics[width=0.9\columnwidth,clip]{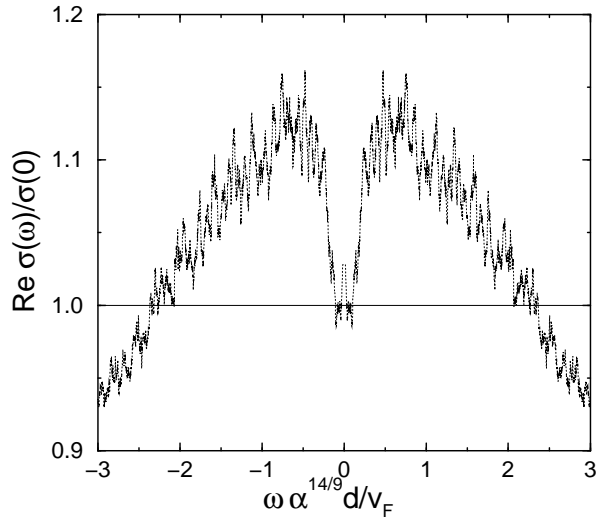}
\vspace{3mm}
\caption{Real part of the {\it ac} conductivity in the RMF at
$\alpha=4.04$. A non-analytic dip around $\omega=0$ is clearly
seen. The low-frequency increase of conductivity is restricted to the
region 
$|\omega|L_s/v_F\lesssim 1$ (where $L_s\sim d\alpha^{14/9}$ is the
length of a link of the percolating network), in agreement with the
theory.}
\label{sxx4}
\end{figure}

Since a real percolating network exhibits all possible sorts of
fluctuations, the fact that we find the $1/t^2$ tail in the model with
fluctuating $\xi_\mu$'s is sufficient to conclude that such a tail
should be present in the problem of the transport in strong RMF.   In
Fig.~\ref{sxx4} we show the results of the numerical simulations of
the problem for $\alpha\simeq 4$. A pronounced dip in the {\it ac}
conductivity around $\omega=0$ in the expected range of frequencies
$|\omega|\lesssim 1/\tau_s\sim v_F/d\alpha^{14/9}$ nicely confirms our
analytical conclusions. The sign of the non-analytic correction
corresponds to a decrease of ${\rm Re}\,\sigma$ as $|\omega|\to 0$.

It is worth mentioning that the problem of a random walk on such a
percolating network is a close relative of the advection-diffusion
problem in a spatially random velocity field ${\bf v}({\bf r})$
(``steady flow") with $\nabla\cdot {\bf v}=0$ (``incompressible
liquid") characterized by the correlation function \begin{equation}
\label{stfl} \int d^2r\exp (-i{\bf qr})\left<v_\alpha (0)v_\beta ({\bf
r})\right>=\tilde{W}_v(q)(\delta_{\alpha\beta}q^2-q_\alpha q_\beta)~.
\end{equation} This model was studied in a series of papers
\cite{addi0,addi1,addi2} with emphasis on the case of long-range
correlations, namely $\tilde{W}_v(q)\propto q^{-2}$ for $q\to 0$
[which corresponds to $\left<v_\alpha (0)v_\beta ({\bf
r})\right>\propto r^{-2}$]. In contrast, we have considered a
percolation lattice with short-scale distortions [$\tilde{W}_v(q)\to
{\rm const}$ at $q\to 0$]. One can check (see Appendix) that the
advection-diffusion problem yields a $t^{-2}$ tail in this case, in
agreement with our consideration above.

\section{Effect of inelastic scattering} 
\label{s3a}

So far our considerations have not included inelastic scattering
processes which change the energy of a particle. The question arises
whether the zero-frequency anomaly $\Delta{\rm Re}\,\sigma(\omega)$ is
cut off at low frequencies $\omega\sim 1/\tau_{\rm in}$, where
$\tau_{\rm in}$ is a relaxation time for the inelastic processes. This
question is studied most conveniently within the Liouville-Boltzmann
approach of Sec.~\ref{s3}. To this end, we consider the
linearized distribution function $\delta f(\omega,{\bf
r},\epsilon,\phi)$ of particles with energy $\epsilon$ and velocity
direction specified by the polar angle $\phi$, subject to a smooth RP
or RMF and inelastic collision processes, obeying the
Liouville-Boltzmann equation
\begin{equation}
\label{inel1}
(-i\omega+{\bf v}{\bf \nabla}+\delta L)\delta f-I_{\rm in}(\delta
f)=S
\end{equation}
with the source term $S=e{\bf v}{\bf E}(\partial f_0/\partial\epsilon)$. 
Here $f_0$ is the Fermi distribution function corresponding to a
temperature $T$ (which we will assume to be low, $T\ll E_F$).  
A simple model form of the collision integral $I_{\rm in}$, which
respects particle number conservation, is sufficient for our purposes:
\begin{equation}
\label{inel2}
I_{\rm in}(\delta f)=-{1\over \tau_{\rm in}}\left[
\delta f(\epsilon,\phi)+{\partial f_0\over\partial \epsilon}\int
d\epsilon'\int {d\phi'\over 2\pi}\delta f(\epsilon',\phi')\right]\ .
\end{equation}
(For simplicity we adopt the model of isotropic, energy independent
inelastic scattering.)

The conductivity is obtained as
\begin{equation}
\label{inel3}
\sigma(\omega)=e^2\nu v_F^2\int d\epsilon \left(-{\partial
f_0\over\partial \epsilon}\right)\int{d\phi\over
2\pi}\left\langle\cos\phi{1\over \tilde{L}_0}\cos\phi\right\rangle\ ,
\end{equation}
where $\tilde{L}_0 = -i\omega+\tau_{\rm in}^{-1}+v_F{\bf n}{\bf
\nabla}$.
Expanding in $\delta L$, averaging over the long-range disorder and
resumming the series, one finds $\sigma(\omega)$ in the form
(\ref{sigmaomega}). The memory function $M$ is now given in lowest
order for the cases of RP and RMF by (\ref{mccv}) and (\ref{mb0}),
respectively, with $\omega$ replaced by $\omega+\tau_{\rm in}^{-1}$. 

In order to calculate the effect of return processes, one needs to
know the diffusion propagator $\tilde{P}_D({\bf
q};\epsilon,\phi;\epsilon',\phi')$ for particles starting with energy
$\epsilon$ and velocity angle $\phi$ and returning with energy
$\epsilon'$ and angle $\phi'$. It obeys Eq.~(\ref{inel1}) with the
source term replaced by
$S_D=\delta(\phi-\phi')\delta(\epsilon-\epsilon')(-\partial
f_0/\partial\epsilon)$. After averaging over $\delta L$, one finds
\begin{eqnarray} \label{inel4} \tilde{P}_D({\bf
q};\epsilon,\phi;\epsilon',\phi')&=&\left({\partial
f_0\over\partial\epsilon}\right){\gamma_\epsilon ({\bf
q},\phi)\gamma_{\epsilon'}({\bf q},\phi')\over
D_tq^2-i\omega}\left({\partial f_0\over\partial\epsilon'}\right)
\nonumber\\ &+& {\rm regular\ terms}\ , \end{eqnarray} where
$D_t=v_F^2\tau_t/2$ with $\tau_t^{-1}=\tau^{-1}+\tau_{\rm in}^{-1}$ is
the total diffusion constant including elastic and inelastic
scattering processes, and $\gamma_\epsilon ({\bf
q},\phi)=[1-iqv(\epsilon)\tau_t\cos(\phi-\phi_q)]$ with
$v(\epsilon)=(2\epsilon /m)^{1/2}$. As expected, the diffusion
propagator shows a diffusion pole even in the presence of inelastic
processes, due to particle number conservation. The scattering ``out"
of particles with given energy $\epsilon$ into other energy states is
exactly compensated by a corresponding scattering-in contribution.

Let us define the function $\Delta M (\epsilon,\epsilon')$ in the same
way as in Eqs.\ (\ref{m2}),(\ref{cmb}) with the only change $P_D({\bf
q},\phi,\phi')\to \tilde{P}_D({\bf
q};\epsilon,\phi;\epsilon',\phi')$. The correction to the memory
function due to return processes, $\Delta M$, is then given by
\begin{eqnarray} \label{inel5} \Delta M=\int d\epsilon d\epsilon'
\Delta M (\epsilon,\epsilon')~.  \end{eqnarray} As a result, the
expressions (\ref{mfrpclass}), (\ref{rpclass}) and (\ref{cmagcon}),
(\ref{cmagres}) remain valid, provided (i) $\tau$ and $l$ are replaced
by the full momentum relaxation time $\tau_t$ and mean free path
$l_t=v_F\tau_t$, respectively, and (ii) additional factors of
$(\tau_t/\tau)^2$ and $\tau_t/\tau$ are included in (\ref{rpclass}),
(\ref{cmagres}), respectively, which stem from the explicit factors of
$\tilde{W}$ in the definition of $\Delta M (\epsilon,\epsilon')$.

Thus, the classical zero-frequency anomaly is not cut off at finite
temperature. This should be contrasted with the quantum zero-frequency
anomaly induced by the weak-localization and Altshuler-Aronov
(interplay of interaction and disorder) effects. It follows that
increasing temperature favors the experimental observation of the
classical anomaly.

\section{Conclusions}
\label{s4}

In this paper, we have studied memory effects in the low-frequency
{\it ac} conductivity of a 2D fermion gas in a long-range random
potential or random magnetic field. We have calculated the long-time
tail in the velocity correlation function induced by diffusive returns
of a particle and leading to a non-analytic $|\omega|$-behavior of the
real part of the conductivity (zero-frequency anomaly). While in a
random potential the $|\omega|$-contribution is positive (as in the
Lorentz gas, Ref.~\cite{ernstwey}) and is proportional to $(d/l)^2$,
where $d/l\ll 1$ is the ratio of the correlation length to the mean
free path, a smooth weak RMF induces a much larger ($\propto d/l$)
correction of opposite sign. The sign difference can be traced back to
the RMF scattering being odd with respect to time reversal.

We have also demonstrated how an $|\omega|$-contribution to ${\rm
Re}\,\sigma(\omega)$ arises in the regime of strong random magnetic
field, where the transport is determined by percolation of the snake
states.  In this case, spatial fluctuations in the geometry of the
percolating network are responsible for the memory effects.

Our numerical simulations confirm the existence of these non-analytic
contributions at low frequency, as well as the unconventional sign of
the correction in the weak random magnetic field. With increasing
strength of the RMF, when the system crosses over into the regime of
the percolating transport, the sign of the effect changes.

The experimental observation of the non-analytic low-frequency
behavior of the {\it ac} conductivity of composite fermions would be
of considerable interest. In particular, we predict that the
$|\omega|$-term in ${\rm Re}\, \sigma$ is negative at $\nu=1/2$ in the
high-mobility samples (where the strength of the effective RMF is
$\alpha\sim 0.3$ \cite{empw}), but should change sign if the system is
driven toward the percolation regime by adding more long-range
scatterers (e.g., antidots \cite{smet97}). A sign change is also
expected with increasing effective magnetic field (moving away from
half-filling), when the system crosses over to a percolation regime,
as discussed in Ref.~\cite{empw}.

\section*{Acknowledgments} We are grateful to D.~Khmelnitskii for
discussions of the role of inelastic scattering and to Y.~Levinson for
attracting our attention to Ref.~\cite{gantsevich81}. This work was
supported by the SFB 195 der Deutschen Forschungsgemeinschaft, by
INTAS grant No. 97-1342, and by the German-Israeli Foundation.

\section*{Appendix}

Consider a particle moving in a diffusive medium with a diffusion
coefficient $D$, subject to a spatially random velocity field ${\bf
v}({\bf r})$. Let the velocity field be incompressible ($\nabla\cdot
{\bf v} =0$) and determined by the correlator (\ref{stfl})
with a finite $\tilde{W}_v(0)$. Assuming the random field to be weak,
we can expand the Green's function
\begin{eqnarray*}
G({\bf r},{\bf r}')=\left<{\bf r}|(-i\omega-D\nabla^2-\nabla\cdot {\bf
v})^{-1}|{\bf r}'\right>
\end{eqnarray*}
in ${\bf v}({\bf r})$, which yields for the Fourier transform
\begin{eqnarray*}
&\tilde{G}&(q)\simeq {1\over -i\omega +Dq^2}\\ \nonumber &+&{1\over
(-i\omega+Dq^2)^2}\left<({\bf v}\cdot\nabla ){1\over -i\omega
-D\nabla^2}({\bf v}\cdot\nabla)\right>_q+\,\dots~.
\end{eqnarray*}
Using the correlation function (\ref{stfl}) and resumming the series,
we thus find
\begin{eqnarray*}
\tilde{G}(q)={1\over -i\omega +(D+\delta D)q^2}~,
\end{eqnarray*}
with the following correction to the diffusion coefficient
\begin{eqnarray*}
\delta D={3\over 2}\int{d^2 q\over (2\pi)^2}{q^2\tilde{W}_v(q)\over
-i\omega +Dq^2}\simeq -{3\over 16}{\tilde{W}_v(0)\over D^2}|\omega|~.
\end{eqnarray*}
Hence, this continuous model predicts a $t^{-2}$ tail in the
velocity-velocity correlation function, with a positive
coefficient. Note, however, that in a lattice model the sign depends on
the microscopic structure of disorder.

\end{multicols}

\begin{references}
\bibitem{forfer} A simple qualitative explanation of the tail
(\ref{lgasv}) is as follows. In the Lorentz gas model, a particle
starting at an arbitrary point at $t=0$ in $x$ direction moves freely
during a time $\sim l/v_F$, until it reaches the first scatterer,
after which it starts to diffuse. The long-time tail arises due to the
memory of the fact that between $t=0$ and the first scattering no
scatterers have been hit, i.e., that the corresponding rectangle with
area $\sim al$ is free of scatterers. The probability of return to
this area is $\sim al/Dt$. Now, returning particles will have on
average a negative projection of $v_x\sim -l/t$. Due to the memory
effect specified above, these particles will not be scattered, thus
giving a negative tail $\left<v_x(t)v_x(0)\right>\sim
-v_F(al/Dt)(l/t)\sim -al/t^2$, in agreement with (\ref{lgasv}). For
the Lorentz gas, the condition for the classical derivation to be
valid is $k_Fa^2\gg l$, where $k_F$ is the Fermi wave vector.
\bibitem{foot2} It is instructive to plot the memory function (rather
than the conductivity), since it would be constant in the Drude
approximation [while ${\rm Re}\,\sigma(\omega)$ has a Lorentzian
shape]. In the weak-disorder regime, deviations from the Drude
approximation are relatively weak and are much better seen in ${\rm
Re}\,\rho(\omega)$ than in ${\rm Re}\,\sigma(\omega)$.
\bibitem{foot3} In a related calculation of the magnetoresistivity
\cite{mwepw} we found that the analytical result for small $\alpha$ is
in quantitative agreement with the numerical simulations for
$\alpha\lesssim 0.01$, while it describes correctly qualitative
features up to $\alpha\sim 0.5$.
\bibitem{aside} The continuum approximation (\ref{adm3}) is not
sufficient to find the numerical prefactor of the correction term in
(\ref{adm4}), i.e., of the contributions $\sim(\tau_s/t)^2$ to
$P_{ij}(t)$.
\end{references}
\end{document}